\documentclass{prop2015v2}
\usepackage[english]{babel}
\title{Model building on the non-factorisable type IIA $\bf{T^6/(\mathbb{Z}_4\times\Omega\mathcal{R})}$ orientifold}
\author[A. Seifert]{Alexander Seifert\inst{1,}\footnote{Corresponding author;\quad 
E-mail:~\textsf{alseifer@uni-mainz.de}}}
\author[S.\,X. Author]{Gabriele Honecker\inst{1}}
\address[1]{Institute for Physics (WA THEP) \&  Cluster of Excellence
PRISMA, Johannes Gutenberg University, D - 55099 Mainz}
\shortauthors{A. Seifert et al.}
\begin{abstract}
We construct  global semi-realistic supersymmetric models with intersecting D6-branes on the non-factorisable orientifold $T^6/(\mathbb{Z}_4\times\Omega\mathcal{R})$. The non-factorisable structure gives rise to differences compared to the factorisable case: additional conditions for the three-cycles to be  Lagrangian and extra constraints on the wrapping numbers for building fractional cycles.
\end{abstract}
\shortabstract
\begin{document}
\maketitle

\section{Geometry of the $T^6/\mathbb{Z}_4$-orbifold}
So far only  factorisable orbifolds has been considered for  model building \cite{Blumenhagen:2002gw} and moduli stabilisation \cite{Ihl:2006pp} on $T^6/(\mathbb{Z}_4 \times \Omega\mathcal{R})$. The models on the non-factorisable ones were to our best knowledge discussed only in \cite{Blumenhagen:2004di} with the restriction that the D6-branes are on the top of  the O6-planes. In our case, we construct  models with  arbitrary D6-branes  on $T^6/(\mathbb{Z}_4 \times \Omega \mathcal{R})$ with the torus lattice $A_3\times A_1\times B_2$, see figure 1.\\ 
The Coxeter element $Q$ acts on this root lattice spanned by the simple roots $\{e_i\}_{i=1,\dots,6}$  as 
\begin{align*}\label{a1}   
&Qe_1=e_2,\quad\; Qe_2:=e_3,\qquad\;\;\;\;Qe_3:=-e_1-e_2-e_3,\\
&Qe_4=-e_4,\;\; Qe_5:=e_5+2e_6,\;\; Qe_6:=-e_5-e_6.
\end{align*}
This action preserves  $\mathcal{N}=2$ supersymmetry in four dimensions and fixes the metric of the six-torus up to four angles and three radii \cite{Lust:2006zh}. The Hodge numbers of this $T^6/\mathbb{Z}_4$-orbifold are  $h_{21}=h_{21}^\text{untw}+h_{21}^{\mathbb{Z}_2}=1+2$ and $h_{11}=h_{11}^\text{untw}+h_{11}^{\mathbb{Z}_4}+h_{11}^{\mathbb{Z}_2}=5+16+6$.
\begin{figure}[h]
	\centering
		\includegraphics[width=8.5cm]{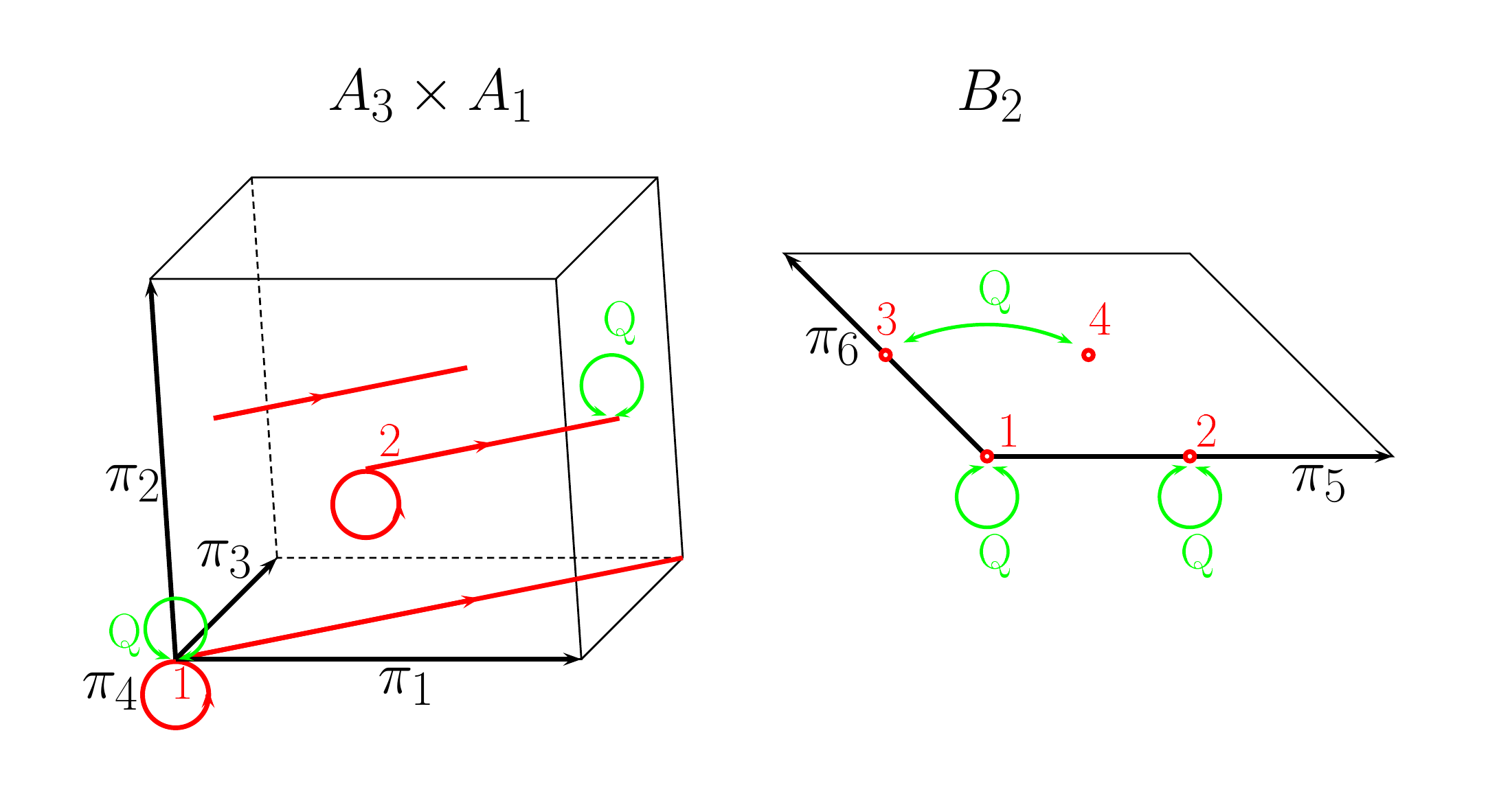}
		\caption{$T^6/\mathbb{Z}_4$-orbifold on the $A_3\times A_1\times B_2$-lattice and its $\mathbb{Z}_2$-fixed lines (in red).}
\end{figure}\\
We first consider the three-cycles inherited from the six-torus and denote  a one-cycle in the direction of the lattice vector $e_i$ by $\pi_i$. The idea is to describe the toroidal three-cycles in a similar way as in the factorisable case, namely by wrapping numbers along the fundamental cycles, so that each three-cycle can be written as a product of three one-cycles. But in contrast to the factorisable case, we also have  cycles like $\pi_{135}$ which wrap a two-cycle on the $A_3$-torus and a one-cycle on the $B_2$-torus.\\
Thus, we can make the ansatz that any toroidal three-cycle  can be described by  ten wrapping numbers s.t. 
\begin{equation}\label{1}
\pi^{\text{torus}}:=\bigwedge_{i=1}^2(m^i\pi_1+n^i\pi_2+p^i\pi_3+q^i\pi_4) \wedge(m^3\pi_5+n^3\pi_6)\,.
\end{equation}
By taking orbits of the $Q$-action $(\tilde{Q}:=\sum_{n=0}^3Q^n)$, a basis of $\mathbb{Z}_4$-invariant {\it bulk} three-cycles is given by
\begin{equation*}
\gamma_1:=-\tilde{Q}\pi_{136}\quad\gamma_2:=-\tilde{Q}\pi_{125},\quad \bar{\gamma}_1:=\tilde{Q}\pi_{146},\quad\bar{\gamma}_2:=\tilde{Q}\pi_{246}.
\end{equation*} 
By means of the $\mathbb{Z}_2$-invariant one-cycles $\pi_1+\pi_3$ and     
$\pi_4$, we can construct the  $\mathbb{Z}_4$-invariant {\it exceptional} three-cycles 
\begin{equation*}
\begin{split}
\gamma_3&:=(e_{13}-e_{14})\wedge(\pi_1+\pi_3)\,,\qquad \bar{\gamma}_3:=(e_{13}-e_{14})\wedge\pi_4\,,\\
\gamma_4&:=(e_{23}-e_{24})\wedge(\pi_1+\pi_3)\,,\qquad \bar{\gamma}_4:=(e_{23}-e_{24})\wedge\pi_4\,,
\end{split}
\end{equation*} 
where $e_{ij}$ are exceptional two-cycles  with the topology of $S^2$ arising from the $\mathbb{Z}_2$-orbifold singularities by blowing up.
The non-vanishing intersection numbers between the bulk and/or exceptional three-cycles are given by
\begin{equation*}
\gamma_i\circ\bar{\gamma}_j=-2\delta_{ij}\quad i,j=1,2,\quad\gamma_m\circ\bar{\gamma}_n=2\delta_{mn}\quad\;\;\;m,n=3,4.
\end{equation*}
Since the intersection form of the $\gamma_i$'s, and $\bar{\gamma}_i$'s $(i=1,2,3,4)$ is not unimodular, these three-cycles do not form an integral basis. Thus, we have to consider fractional three-cycles, which can consist of  half a bulk cycle and simultaneously of half an exceptional cycle.\\
If in the case of the factorisable torus  every bulk cycle passing through  $\mathbb{Z}_2$-invariant points in $(T^2)^3$ can split into fractional cycles,  here  this does not hold true any more, so that we have to consider  
\begin{equation*}
\pi\text{ is fractional }\Longleftrightarrow\;Q^2\pi^{\text{torus}}=\pi^{\text{torus}}\,,
\end{equation*}
which leads to some constraints on the wrapping numbers $(m^i,n^i,p^i,q^i)$. Furthermore, we find that the fractional cycles form the $F_4\oplus F_4$ lattice, the basis of which gives rise to the unimodular basis of $H_3(T^6/\mathbb{Z}_4;\mathbb{Z})$.

\section{IIA Orientifolds on the $\mathbb{Z}_4$-orbifold}
\subsection{Supersymmetry}
The orientifold projection $\Omega\mathcal{R}$, where $\mathcal{R}$ is an anti-holomorphic involution $z_i\rightarrow e^{i\theta_i}\bar{z}_i$, breaks supersymmetry  to $\mathcal{N}=1$.
The O6-planes are given by the fixed point loci  of the involution.
Their presence  enables the fulfilment of the RR-tadpole cancellation conditions, which allows for the existence of  global models.\\
The  $T^6/\mathbb{Z}_4$-orbifold has only one continuous complex structure $\mathcal{U}$  inherited from the six-torus~\cite{Lust:2006zh}. 
The crystallographic action of $\mathcal{R}$  on the root lattice then fixes  Re$(\mathcal{U})=0,\frac{1}{2}$ (for  $\theta_i=0$).\\
We are interested in supersymmetric D6-branes. This implies that all three-cycles wrapped by them have to be  \textit{special Lagrangian} ({\it sLag}):
\begin{equation*}
J\mid_{\pi^{\text{torus}}}=0\,,\quad
\text{Im}(\Omega_3)\mid_{\pi^{\text{torus}}}=0\,,\quad
\text{Re}(\Omega_3)\mid_{\pi^{\text{torus}}}>0\,,
\end{equation*}
where   $\Omega_3$ is the holomorphic volume-form and  $J$ the K\"ahler-form.
The anti-symmetry of the K\"ahler form under the involution $(\mathcal{R}J=-J)$ fixes one of the angle-moduli.
Contrary to the factorisable case, where the three-cycles satisfy the {\it Lag} condition $(J\mid_{\pi^{\text{torus}}}=0)$ automatically, in the present non-factorisable case the {\it Lag} condition gives rise to four constraints on the wrapping numbers of the toroidal three-cycle of eq \eqref{1}. Half of them can be fulfilled by fixing an angular modulus and the remaining constraints are equivalent to the fractional ones, i.e.\;any {\it Lag} three-cycle passing through the $\mathbb{Z}_2$-fixed points is $\mathbb{Z}_2$-invariant and thus can be made fractional, and vice versa.
\subsection{Supersymmetric Pati-Salam models}
The involution $\mathcal{R}$ (for $\theta_i=0$)  gives rise to  two possible lattice configurations with Re$(\mathcal{U})=0$ or $\frac{1}{2}$.
For the lattice with  Re$(\mathcal{U})=\frac{1}{2}$, there exist some {\it local} Pati-Salam (PS) models with three generations, but these come with chiral matter states transforming in the anti-symmetric representation of the gauge group of the $a$-stack.
We also found {\it global} PS models with two and four generations for both lattices. The examplary chiral spectrum of a {\it global} 
four-generation PS model is displayed in the table below:\\ 
\begin{tabular}{|m{1.3cm}|p{6.49cm}|}
\hline 
\multicolumn{2}{|m{8cm}|}{Chiral spectrum of a global Pati-Salam model
 with four generations (for Re$(\mathcal{U})=0$)}\\\hline
sector & $SU(4)_a\times [USp\,\text{or}\,SO](2)_b\times [USp\,\text{or}\,SO](2)_c\times [USp\;\text{or}\;SO](4)_d\times U(1)_a$\\ 
\hline 
$ab=ab^\prime$ & $4\times(4,\bar{2},1,1)_{+1}$ \\ 
\hline 
$ac=ac^\prime$ &$4\times(\bar{4},1,2,1)_{-1}$ \\ 
\hline 
\end{tabular}
\section{Results and Outlook}
We studied the geometry of the non-factorisable $T^6/(\mathbb{Z}_4\times \Omega\mathcal{R})$-orientifold and verified the analogies and differences to the factorisable case. We found that global supersymmetric  PS models with two and four generations are possible. 
We aim at extending the search for global three-generation PS models by the analysis of different choices of the involution $\mathcal{R}$,
and we will classify the type of symmetry enhancement [$USp\;\text{or}\;SO$] based on conformal field theory methods, see e.g.~\cite{Honecker:2011sm}.
Last but not least, studying non-factorisable tori is motivated by moduli stabilisation involving closed string background fluxes, which we plan to address in the future.
\begin{acknowledgements}
This work is partially supported by
the Cluster of Excellence PRISMA DGF no. EXC 1098, the
DFG research grant HO 4166/2-1 and the GRK 1581.
\end{acknowledgements}

\end{document}